\documentclass[preprint]{revtex4}
\usepackage{graphicx}
\usepackage{bm}
\usepackage{psfrag}
\usepackage{amsmath}
\usepackage{amsfonts}
\usepackage{amssymb}
\usepackage[latin1]{inputenc}

\newcommand{\ket}[1]{|#1\rangle}
\newcommand{\bra}[1]{\langle#1|}

\newcommand{\scalar}[2]{\langle#1|#2\rangle}

\newcommand{\op}[1]{|#1\rangle\langle#1|}

\newcommand{\spinor}[2]{\left(\begin{array}{c}#1\\#2\end{array}\right)}
\newcommand{\cn}{\frac{1}{\sqrt{2}}}

\newcommand{\intk}{\int_{-\pi}^\pi \frac{dk}{2\pi}}

\newcommand{\real}[1]{\mbox{Re}\left[#1\right]}
\begin{document}

\title{Quantum walk on the line: \\entanglement and non-local initial conditions}
\author{G. Abal}
\email{abal@fing.edu.uy}
\author{R. Siri}
\author{A. Romanelli}
\author{R. Donangelo}
\altaffiliation[Permanent address: ]
{Instituto de F\'{\i}sica, Universidade Federal do Rio de Janeiro, C.P. 68528, Rio de Janeiro 21941-972, Brazil}
\affiliation{Instituto de Física, Universidad de la República \\ 
C.C. 30, C.P. 11000, Montevideo, Uruguay
}
\date{\today}
\begin{abstract}
The conditional shift in the evolution operator of a quantum walk generates entanglement between the coin and position degrees of freedom. This entanglement can be quantified by the von Neumann entropy of the reduced density operator (entropy of entanglement). In the long time limit, it converges to a well defined value which depends on the initial state. Exact expressions for the asymptotic (long-time) entanglement are obtained for (i) localized initial conditions and (ii) initial conditions in the position subspace spanned by $\ket{\pm 1}$.  
\end{abstract}



\maketitle

\section{Introduction}
\label{sec:intro}

Quantum walks in several topologies \cite{Kempe03} are being studied as potential sources for new quantum algorithms. Recently, quantum search algorithms based on different versions of the quantum walk, have been proposed \cite{Shenvi,Childs}. These algorithms take advantage of quantum parallelism, but do not rely on entanglement, which has only recently begun to be addressed in the context of quantum walks. The first studies \cite {Venegas,Omar} where numerical and considered walkers driven by two coins which where maximally entangled by their initial condition. More recently,  the coin-position entanglement induced by the the evolution operator of a quantum walk on the line was investigated numerically. For pure states, entanglement can be quatified using the von Neumann entropy of the reduced density operator (entropy of entanglement). Using this measure, it was suggested  \cite{Carneiro} that for all coin initial states of a Hadamard walk the entanglement would have a long-time value close to $0.872$. In this work, we use the Fourier representation of the Hadamard walk on a line to investigate the dependence of asymptotic entanglement with the initial conditions. This powerful approach enables us to derive the condition that must be satisfied by the initial state in order to obtain a limiting value of $0.872$. However, for localized initial states with generic coins the asymptotic entanglement changes smoothly between $\sim 0.736$ and $1$ (maximum entanglement).  Furthermore, if non-local initial conditions are considered, all entanglement levels are accessible in the long-time limit. We consider in detail the case of non-local initial conditions restricted to a certain position subspace. 

This work is organized as follows. In Section~\ref{sec:qwalk}, we briefly review the discrete-time quantum walk on the line and define the entropy of entanglement. To illustrate the general method, the asymptotic value for the entanglement of a particular localized initial condition is obtained analytically. An expression for the dependence of the asymtptotic entanglement for generic localized initial conditions is derived in the Appendix. Section~\ref{sec:non-local} is devoted to the calculation of the asymptotic entanglement induced by the evolution operator of a Hadamard walk for a particular class of non-local initial conditions.  Finally, in Section~\ref{sec:conc} we summarize our conclusions and discuss future developments. 

\section{Discrete-time quantum walk on the line}
\label{sec:qwalk}

The discrete-time quantum walk can be thought as a quantum analog of the classical random walk where the classical coin flipping is replaced by a Hadamard operation in an abstract two-state quantum space (the coin space). A step of the quantum walker consists of a conditional traslation on the line. Of course, if the quantum coin is measured before taking a step, the classical walk is recovered \cite{qw-markov,deco}. The Hilbert space ${\cal H}={\cal H}_P\otimes{\cal H}_C$, is composed of two parts: a spatial subspace, ${\cal H}_P $, spanned by the orthonormal set $\{\ket{x}\}$ where the integers $x=0,\pm 1,\pm 2\ldots$ are associated to discrete positions on the line and a single-qubit coin space, ${\cal H}_C$,  spanned by two orthonormal vectors denoted $\{\ket{R}, \ket{L}\}$. A generic state for the walker is 
\begin{equation}
\ket{\Psi}=\sum_{x=-\infty}^\infty \ket{x}\otimes\left[ a_x\ket{R} + b_x\ket{L}\right] \label{sp-wv}
\end{equation} 
in terms of complex coefficients satisfying the normalization condition $\sum_x |a_x^2|+|b_x|^2=1$ (in what follows, the summation limits are left implicit). 

A step of the walk is described by the unitary operator
\begin{equation}
U=S\cdot\left(I_P\otimes C\right) \label{evol1}
\end{equation}
where $C$ is a suitable unitary operation in ${\cal H}_C$ and $I_P$ is the identity in ${\cal H}_P$. A convenient choice is a Hadamard operation, $\protect{H\ket{R}=\left(\ket{R}+\ket{L} \right)/\sqrt{2}}$ and $\protect{H\ket{L}=\left(\ket{R}-\ket{L} \right)/\sqrt{2}}$. 
When $C=H$, as in this work, one refers to the process as a Hadamard walk. The shift operator 
\begin{equation}
S=S_R\otimes\op{R} + S_L\otimes\op{L} \label{shift-x}
\end{equation} 
with $\protect{S_R=\sum_x \ket{x+1}\bra{x}}$ and $\protect{S_L=S_R^\dagger=\sum_x \ket{x-1}\bra{x}}$, conditionally shifts the position one step to the right (left) for coin state R (L). Furthermore, it generates entanglement between the coin and position degrees of freedom. 

The evolution of an initial state $\ket{\Psi(0)}$ is given by 
\begin{equation}
\ket{\Psi(t)}=U^t\ket{\Psi(0)}\label{1p-map}
\end{equation} 
where the non-negative integer $t$ counts the discrete time steps that have been taken. The probability distribution for finding the walker at site $x$ at time $t$ is $\protect{P(x; t)=|\scalar{x}{\Psi(t)}|^2=|a_x|^2+|b_x|^2}$. 
The variance of this distribution  increases quadratically with time \cite{Travaglione} as opposed to the classical random walk, in which the increase is only linear.  This advantage in the spreading speed in the quantum case is directly related to quantum interference effects and is eventually lost in the presence of decoherence \cite{deco}. 

One of the early papers on quantum walks, due to Nayak and Vishwanath \cite{Nayak}, has shown that Fourier analysis can be successfully used to obtain integral expressions for the amplitudes $a_x(t)$ and $b_x(t)$ for given initial conditions. The resulting quadratures can be evaluated in the long-time limit. The usefulness of this approach has been limited because the detailed calculations are lengthy and must be individually worked out for each initial condition. In this paper, we shall use the dual Fourier space to obtain information about the asymptotic entanglement, and so we review those results. 
 
\subsection*{Fourier transform}

The dual space $\tilde{\cal H}_k$ is spanned by the Fourier transformed kets $\ket{k}=\sum_x e^{ikx}\ket{x}$, where the wavenumber $k$ is real and restricted to $[-\pi,\pi]$. The state vector (\ref{sp-wv}) can then be written 
\begin{equation}
\ket{\Psi}=\intk\ket{k}\otimes\left[\tilde a_k\ket{R} + \tilde b_k\ket{L}\right]\label{wv-k}
\end{equation} 
where the k-amplitudes $\tilde a_k=\scalar{k,R}{\Psi}$ and $\tilde b_k=\scalar{k,L}{\Psi}$ are related to the position amplitudes by 
\begin{equation}
\tilde a_k = \sum_x e^{-ikx}a_x\qquad\mbox{and}\qquad\tilde b_k = \sum_x e^{-ikx}b_x.\label{ft-a}
\end{equation}
The shift operator, defined in (\ref{shift-x}), is diagonal in $\tilde{\cal H}_k$ space: $\protect{S\ket{k,R}=e^{-ik}\ket{k,R}}$ and $\protect{S\ket{k,L}=e^{ik}\ket{k,L}}$. A step in the evolution may be expressed through the evolution operator $U_k$ in $k$--space as 
\begin{equation}
\ket{\Phi_k(t+1)}=U_k\ket{\Phi_k(t)}=\frac{1}{\sqrt{2}}\left(
\begin{array}{cc}
e^{-ik} & e^{-ik} \\ 
e^{ik} & -e^{ik}
\end{array} \right)
\ket{\Phi_k(t)}\label{k-evol}
\end{equation} 
where $\ket{\Phi_k}=\scalar{k}{\Psi}$ is the spinor $(\tilde a_k, \tilde b_k)^T$.
This operator has eigenvectors $\ket{\varphi_k^{(1,2)}}$ given by 
\begin{eqnarray}
\ket{\varphi_k^{(1)}}=\alpha_k\spinor{u_k}{v_k}&\qquad&\ket{\varphi_k^{(2)}}=\beta_k\spinor{u_k}{w_k}\label{eval}
\end{eqnarray} 
where $\alpha_k$ and $\beta_k$ are the real, positive functions, 
\begin{eqnarray}
\alpha_k &\equiv&\cn\left[1+\cos^2k-\cos k\sqrt{1+\cos^2k}\right]^{-1/2}\nonumber\\
\beta_k &\equiv&\cn\left[1+\cos^2k+\cos k\sqrt{1+\cos^2k}\right]^{-1/2}\label{alpha-beta}
\end{eqnarray} 
and 
\begin{eqnarray}
u_k&\equiv& e^{-ik}\nonumber\\
v_k&=&\sqrt{2}e^{-i\omega_k}-e^{-ik}\label{ev-part}\\
w_k&=&-\sqrt{2}e^{i\omega_k}-e^{-ik}.\nonumber
\end{eqnarray}
The frequency $\omega_k$, defined by 
\begin{equation}
\sin\omega_k\equiv\frac{\sin k}{\sqrt{2}}\qquad\omega_k\in[-\pi/2,\pi/2],
\label{wk}
\end{equation} 
determines the eigenvalues, $\pm e^{\mp i\omega_k}$, of $U_k$. Using the spectral decomposition for $U_k$, the time evolution of an initial spinor can be expressed as
\begin{eqnarray}
\ket{\Phi_k(t)}&=&U_k^t\ket{\Phi_k(0)}=\label{k-evol1}\\
&&\quad e^{-i\omega_k t}\scalar{\varphi_k^{(1)}}{\Phi_k(0)}\;\ket{\varphi_k^{(1)}}\nonumber\\
&&\qquad +(-1)^t e^{i\omega_k t}\scalar{\varphi_k^{(2)}}{\Phi_k(0)}\;\ket{\varphi_k^{(2)}}.\nonumber
\end{eqnarray} 
In principle, this expression can be transformed back to position space and the probability distribution $P(x,t)$ can be obtained. This approach requires evaluation of complicated integrals which, for arbitrary times, can only be done numerically. However, in the long time limit, stationary phase methods can be used  to approximate the resulting integrals for given initial conditions, an approach illustrated in  \cite{Nayak}. In this work, we bypass these technical difficulties, because  the asymptotic entanglement introduced by $U_k^t$, may be quantified directly from eq.~(\ref{k-evol1}), without transforming back to position space. 

\subsection*{Entropy of entanglement}

Consider the density operator $\rho=\op{\Psi}$. 
Entanglement for pure states can be quantified by the von Neumann entropy of the reduced density operator $\rho_c=tr(\rho)$, where the partial trace is taken over position (or alternatively,  wavenumber $k$). Note that, in general $\protect{tr(\rho_c^2)<1}$, i.e. the reduced operator $\rho_c$ corresponds to a statistical mixture. The associated von Neumann entropy 
\begin{equation}
S_E=-tr(\rho_c\log_2\rho_c),\label{SC}
\end{equation} 
also known as entropy of entanglement, quantifies the quantum correlations present in the pure state $\rho$ \cite{Myhr}. It is zero for a product state and unity for a maximally entangled coin state. It is also invariant under local unitary transformations, a usual requirement for entanglement measures \cite{Vedral,SM95}. 

The entropy of entanglement can be obtained after diagonalisation of $\rho_c$. 
This operator, which acts in ${\cal H}_C$, is represented by the Hermitean matrix
\begin{equation}
\rho_c=\left(
\begin{array}{cc}
 A& B \\ 
 B^*&C 
\end{array}\right), \label{rho_C}
\end{equation} 
where 
\begin{eqnarray}
A&\equiv&\sum_x|a_x|^2=\intk |\tilde a_k|^2\nonumber\\
B&\equiv&\sum_x a_xb_x^*=\intk \tilde a_k\tilde b_k^*\label{coefs}\\
C&\equiv&\sum_x|b_x|^2=\intk |\tilde b_k|^2.\nonumber
\end{eqnarray} 
Normalization requires that tr$(\rho)=1$ and $A+C=1$. In terms of 
\begin{equation}
\Delta=AC-|B|^2\label{disc}
\end{equation}
the real, positive eigenvalues $r_{1,2}$ of this operator are given by 
\begin{equation}
r_{1,2}=\frac12\left[1\pm\sqrt{1-4\Delta} \right] \label{rho-ev}
\end{equation} 
and the reduced entropy can be calculated from eq.~(\ref{SC}) as 
$S_E=-\left( r_1\log_2r_1 + r_2\log_2r_2\right) $. In the rest of this work, we shall be concerned with clarifying the asymptotic (long time) value of $S_E$ for both local and non-local initial conditions. 

\subsection*{Asymptotic entanglement from local initial conditions: a simple example}
\label{ssec:local-ex}
As a simple application, consider the particular localized initial condition 
\begin{equation}
\ket{\Psi(0)}=\ket{0}\otimes\ket{L}, \label{0L}
\end{equation} 
which implies $a_k(0)=0$ and $b_k(0)=1$. For this simple case, the spinor components at time $t$ from eq.~(\ref{k-evol1}), have been explicitly calculated in Ref.~\cite{Nayak} as
\begin{eqnarray}
\tilde a_k(t)&=&
\frac{i e^{ik}}{2\sqrt{1+\cos^2k}}
\left(e^{-i\omega_k t}-(-1)^t e^{i\omega_k t}\right)\nonumber\\
\tilde b_k(t)&=&\frac12\left(1+\frac{\cos k}{\sqrt{1+\cos^2k}}\right)e^{-i\omega_k t}\label{nayak}\\
&&\quad+\frac{(-1)^t}{2}\left(1-\frac{\cos k}{\sqrt{1+\cos^2k}}\right)e^{i\omega_k t}.\nonumber
\end{eqnarray} 

The relevant quantities for the entropy entanglement are $A$ and $B$, defined in eqs.(\ref{coefs}). After some manipulation, from (\ref{nayak}) we obtain
\begin{eqnarray}
A(t)&=&\intk 
\frac{1-(-1)^t\cos(2\omega_k t)}{2(1+\cos^2 k)}\nonumber\\
B(t)&=&\intk 
\frac{ie^{ik}}{2\sqrt{1+\cos^2 k}}\times\label{a2-nayak}\\
&&\left[1-(-1)^t\left(\frac{\cos(2\omega_k t)\cos k}{\sqrt{1+\cos^2 k}}+
 i\sin(2\omega_k t)\right)\right].\nonumber
\end{eqnarray} 
The time dependence of these expressions vanishes in the long time limit and we obtain the exact asymptotic values, 
\begin{eqnarray}
\bar A&=&\lim_{t\rightarrow\infty}A(t)=\frac12\intk\;\frac{1}{1+\cos^2k}=\frac{\sqrt{2}}{4}\\
\bar B&=&\lim_{t\rightarrow\infty}B(t)=\frac{i}{2}\intk\;\frac{\cos^2 k}{1+\cos^2k}=i\frac{2-\sqrt{2}}{4}.\nonumber
\end{eqnarray}  
From eq.~(\ref{disc}) we obtain 
\begin{equation}
\Delta=\Delta_0=\frac{\sqrt{2}}{2}-\frac 12\label{disc0}
\end{equation} 
and the exact eigenvalues of the reduced density operator are $r_1=1/\sqrt{2}$ and $1-1/\sqrt{2}$. 
These eigenvalues yield the asymptotic value for the entropy of entanglement $\bar S_0\approx 0.87243\ldots$, in agreement with the numerical observations reported in Ref.~\cite{Carneiro}. When the procedure outlined above is repeated for arbitrary initial coin states, the asymptotic entanglement level is dependent on the initial coin state. Let us parametrize the initial coin in terms of two real angles $\alpha\in [-\pi,\pi]$ and $\beta\in[-\pi,\pi]$, 
\begin{equation}
\ket{\chi}=\frac{1}{\sqrt{2}}\left( \cos\alpha\ket{R}+e^{i\beta}\sin\alpha\ket{L}\right) .\label{ic1}
\end{equation} 
\begin{figure}
\begin{center}
\includegraphics[scale=0.5,angle=-90]{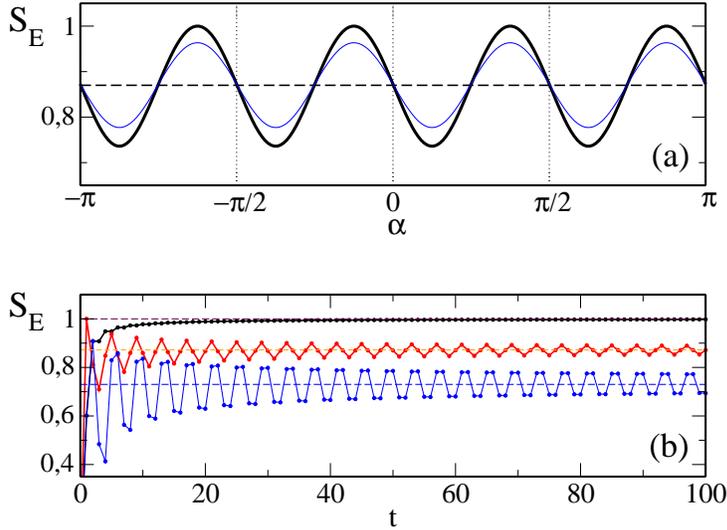}
\end{center}
\caption{\small Panel (a):  asymptotic entropy of entanglement, $\bar S_E$, for localized initial states as a function of $\alpha$ and $\beta$, from  eq.~(\ref{ic1}) (color online). For $\beta=0$, (thick line), $\beta=\pi/4$, (thin line) and $\beta=\pi/2$ (dashed line).  Panel (b): time evolution of the entropy of entanglement, $S_E$, for $\alpha=-\pi/8$ and $\beta=0, $ (full asymptotic entanglement), $\beta=\pi/2$ (intermediate asymptotic entanglement) and  $\beta=\pi/4$ (minimimum asymptotic entanglement). }
\label{fig:SE_local}
\end{figure}
Then, the dependence of $\Delta$, defined in eq~(\ref{disc}), on the initial coin is 
\begin{equation}
\Delta=\Delta_0-2b_1^2\cos\beta\sin(4\alpha),\label{disc2}
\end{equation} 
with $\Delta_0$ given in eq.~(\ref{disc0}) and $b_1=(2-\sqrt{2})/4$. 
We include the details of the calculation leading to this expression in Appendix A.  Thus, $\Delta(\alpha,\beta)$ is a smooth, odd, periodic function of $\alpha$ and only in the special cases \newpage
\begin{enumerate}
\item[i)] $\beta=\pm\pi/2$; for arbitrary $\alpha$
 \item[ii)] $\alpha=0,\pm\pi/4,\pm\pi/2,\pm 3\pi/4$,$\pm\pi$; for arbitrary $\beta$
\end{enumerate}
the asymptotic entanglement is $S_E\approx 0.872\ldots$. In panel (a) of Fig.~\ref{fig:SE_local} the variation of the asymptotic entanglement associated to localized initial coins is shown. Full entanglement can be obtained for special initial coins. In panel (b) the time evolution of the entropy of entaglement is shown in three special cases with $\alpha=-3\pi/8$: for $\beta=0$, full entanglement is achieved rather fast and the oscillations are supressed. For $\beta=\pi/2$ the intermediate asymptotic entanglement level $0.872\ldots$ results. For $\beta=\pi$, the minimum asymptotic entanglement is obtained, but the oscillations are larger and the convergence slow. If non-local initial conditions are considered, lower values for asymptotic entanglement may be obtained.

\section{Non-local initial conditions}
\label{sec:non-local}
\begin{figure}
\includegraphics[width=8cm]{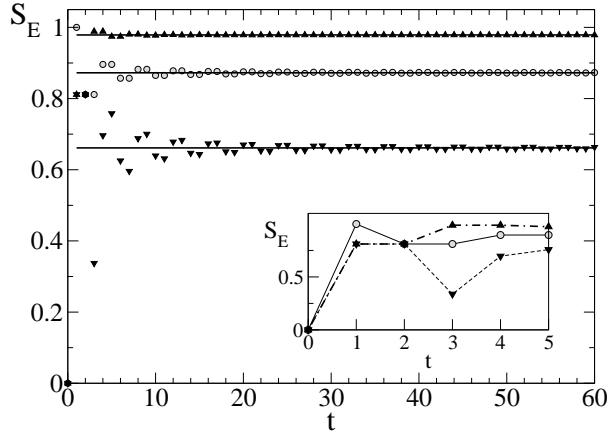}
\caption{Evolution of the entropy of entanglement for the initial delocalized sates from eq.~(\ref{Psi_s}) (up and down black triangles respectively) and for a localized state $\ket{0}\otimes\ket{\chi}$ (gray circles). The time evolution was calculated from eqs.(\ref{a2}). The horizontal lines represent the asymptotic entanglement levels, $0.661$, $0.872$ and $0.979$, obtained  from eqs.(\ref{ab11}) as explained in the text. The inset shows the first steps in detail. } 
\label{fig:ent}
\end{figure}

Most previous work on quantum walks has dealt with initial wavevectors localized in a position eigenstate $\ket{0}$.  When non-local initial conditions are considered, new features emerge. Let us consider a quantum walk initialized in a simple uniform superposition of two position eigenstates such as, 
\begin{equation}
\ket{\Psi_\pm}=\frac{\ket{-1}\pm \ket{+1}}{\sqrt{2}}\otimes\ket{\chi}
\label{Psi_s}
\end{equation}
with the initial coin fixed at $\ket{\chi}=(\ket{R}+i\ket{L})/\sqrt{2}$. For this initial coin, any localized initial condition yields asymptotic entanglement $\bar S_E\approx 0.872\ldots$.  The entanglement induced by the evolution operator when starting from these initial states is shown in Fig.~\ref{fig:ent}. The asymptotic values are $\protect{\bar S_+\approx 0.979}$  and $\protect{\bar S_-\approx 0.661}$, respectively. Below, we provide an analytical explanation for these observed values.

As can be seen in Fig.~\ref{fig:ent}, the rate at which the asymptotic value for entanglement is approached is faster for higher asymptotic entanglement levels. The inset in this figure shows the first five steps in detail. The local initial condition $\ket{\Psi_0}$ is fully entangled $(S_E=1)$ after the first time step and reaches its asymptotic level after $\sim 10$ steps. The non local conditions $\ket{\Psi_\pm}$ have the same evolution for $S_E$ in the first two steps, but the phase difference causes very different entanglement levels after the third time step:  $\ket{\Psi_+}$ reaches its asymptotic level after three steps, while $\ket{\Psi_-}$ takes about 30 time steps to stabilize.

\subsection*{Asymptotic entanglement}
\label{ssec:asymp}

Consider the problem of determining the asymptotic entanglement for non-local initial conditions of the form 
\begin{equation}
\ket{\Psi(0)}=\sum_x c_x(0)\ket{x}\otimes\ket{\chi}\label{Psi_x}
\end{equation} 
with real $c_x(0)$. The coin state $\ket{\chi}$  in eq.~(\ref{Psi_x}) is such that 
\begin{equation}
b_x(0)=ia_x(0)=\frac{c_x(0)}{\sqrt{2}}~\mbox{ or, equivalently,}~\tilde b_k(0)=i\tilde a_k(0).\label{sym}
\end{equation} 
This restriction considerably simplifies the algebra, but it is not essential and the method applies equally well to arbitrary initial coin states. 

\begin{figure}
\begin{center}
\includegraphics[scale=0.5]{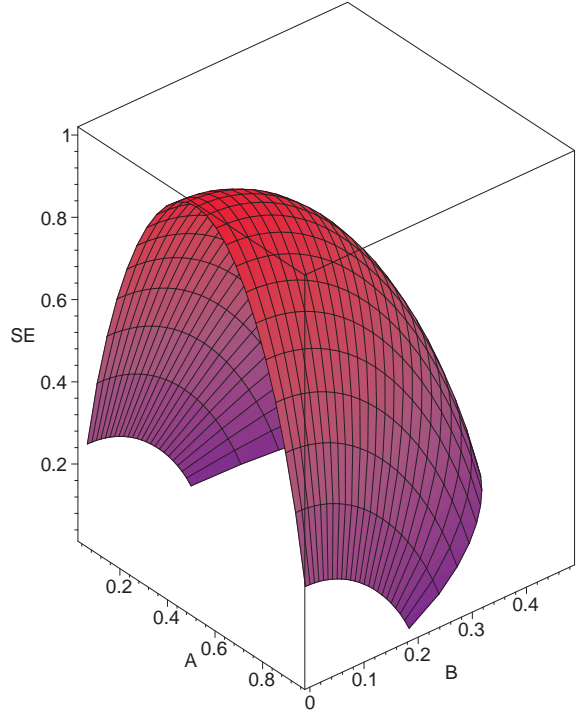}
\end{center}
\caption{\small Asymptotic entropy of entanglement, $\bar S_E$, as a function of $A$ and $|B|$, defined in eqs.(\ref{coefs})} 
\label{fig:sc-ab}
\end{figure}

The eigenvalues of the reduced density operator depend on the real coefficients $A,C$ and on the complex one $B$, defined in eqs.(\ref{coefs}). Fig.~\ref{fig:sc-ab} shows the entropy of entanglement as a function of these coefficients. Maximum entanglement would be obtained for $A=1/2$ and $B=0$, when the reduced operator corresponds to the minimum information mixture $\rho_c=I/2$. 

In order to find expressions for $\bar A$ and $\bar B$, we start by rewriting eq.~(\ref{k-evol1}) in the more explicit form,
\begin{eqnarray}
\tilde a_k(t)=\alpha_k^2 F_k u_k e^{-i\omega_k t}+(-1)^t \beta_k^2 G_k u_k e^{i\omega_k t}\nonumber\\
\tilde b_k(t)=\alpha_k^2 F_k v_k e^{-i\omega_k t}+(-1)^t \beta_k^2 G_k w_k e^{i\omega_k t}.\label{amps}
\end{eqnarray} 
Here, $\alpha_k$ and $\beta_k$ are the real, positive functions defined in eq.~(\ref{alpha-beta}) and $u_k,v_k$ and $w_k$, the other part of the eigenvectors of $U_k$, are defined in (\ref{ev-part}). The dependence on the initial conditions is contained in the complex factors
\begin{eqnarray}
F_k\equiv u_k^*\tilde a_k(0) + v_k^* \tilde b_k(0)\nonumber\\
G_k\equiv u_k^*\tilde a_k(0) + w_k^* \tilde b_k(0).\label{fg-exps}
\end{eqnarray} 


The required expressions for $|\tilde a_k|^2, |\tilde b_k|^2$ and $\tilde a_k\tilde b_k^*$ are 
\begin{eqnarray}
|\tilde a_k(t)|^2&=&\alpha_k^4|F_k|^2+\beta_k^4|G_k|^2 +(-1)^t 2\alpha_k^2\beta_k^2\real{F_kG_k^*e^{-2i\omega_k t}}\nonumber\\
|\tilde b_k(t)|^2&=&\alpha_k^4|F_k|^2|v_k|^2+\beta_k^4|G_k|^2|w_k|^2
+(-1)^t 2\alpha_k^2\beta_k^2\real{F_kG_k^*v_kw_k^*e^{-2i\omega_k t}}\label{a2}\\
\tilde a_k(t)\tilde b_k^*(t)&=&\alpha_k^4|F_k|^2 u_kv_k^*+\beta_k^4|G_k|^2u_kw_k^*+(-1)^t\alpha_k^2\beta_k^2\left[F_kG_k^*u_kw_k^*e^{-2i\omega_k t}
+F_k^*G_k u_kv_k^*e^{2i\omega_k t}\right].\nonumber
\end{eqnarray} 

In the long--time limit, the contribution of the time-dependent terms in the $k-$integrals of eqs.(\ref{a2}) vanishes as $t^{-1/2}$, as shown in detail in \cite{Nayak}. The asymptotic values $\bar A, \bar B$ and $\bar C$ can be obtained from the time--independent expressions, 
\begin{eqnarray}
\bar A &=&\intk \left( \alpha_k^4|F_k|^2+\beta_k^4|G_k|^2\right) \nonumber\\
\bar C &=&\intk \left( \alpha_k^4|F_k|^2|v_k|^2+\beta_k^4|G_k|^2|w_k|^2\right) \label{ab11}\\
\bar B &=&\intk \left( \alpha_k^4|F_k|^2u_kv_k^* +\beta_k^4|G_k|^2u_kw_k^*\right) ,\nonumber
\end{eqnarray} 
which hold for arbitrary initial conditions. 

Now we particularize these expressions for the initial states defined in eq.~(\ref{Psi_s}). Using the symmetry condition, eq. (\ref{sym}), the required squared moduli $|F_k|^2$ and $|G_k|^2$ can be expressed as 
\begin{eqnarray}
|F_k|^2&=&4|\tilde a_k(0)|^2\left[1-\cos(k-\omega_k+\pi/4)\right]\nonumber\\
|G_k|^2&=&4|\tilde a_k(0)|^2\left[1+\cos(k+\omega_k+\pi/4)\right].\label{gk}
\end{eqnarray} 

Since $\bar A+\bar C=1$, we need only perform the integration for $\bar A$,  
\begin{equation}
\bar A=\intk |\bar a_k|^2 = \intk Q(k) |\tilde a_k(0)|^2 \label{b2-ave}
\end{equation} 
where $Q(k)$  is a real function, 
\begin{equation}
Q(k)=4\left[\alpha_k^4\left( 1-\cos(k-\omega_k+\pi/4)\right) +\beta_k^4\left(1+\cos(k+\omega_k+\pi/4)\right)\right]\label{Q-fun0}
\end{equation}
which, after some manipulation, can be simply expressed as 
\begin{equation}
Q(k)= 1+\frac{\sin k\cos k}{1+\cos^2k}.\label{Q-fun}
\end{equation} 
Note that this weight function satisfies $\protect{Q(k)>0}$ and $\protect{\intk Q(k)=1}$.

The asymptotic form for $\protect{\bar{B}}$ is obtained from 
\begin{equation}
\bar B=\intk\,\overline{a_kb_k^*}\equiv \intk R(k)~|\tilde a_k(0)|^2 \label{b-abs}
\end{equation} 
whith $R(k)$ given by,  
\begin{equation}
R(k)= \frac{e^{-ik}\sin k}{1+\cos^2 k}.\label{Rk}
\end{equation}


The required quantity, $\bar S_E$, can now be evaluated  from eqs.(\ref{b2-ave}) and (\ref{b-abs}) for different initial conditions satisfying eq.~(\ref{sym}).

\label{ssec:local}
For the local initial condition $\ket{0}\otimes\ket{\chi}$,  we have $|\tilde a_k(0)|^2=1/2$ and, from eq.~(\ref{b2-ave}), $\bar A=\bar C=1/2$ results immediately. In this case, eq.~(\ref{rho-ev}) reduces to $\bar r_{1,2}=\frac12 \pm |\bar B|$ and the asymptotic eigenvalues are determined by $|\bar B|$ alone. This quantity is obtained from eq.~(\ref{b-abs}), after noticing that if $|\tilde a_k(0)|^2$ is an even function of $k$, only the imaginary part of $R(k)$ contributes. We obtain, 
\begin{equation}
B_0=|\bar{B}(\Psi_0)|=\frac{\sqrt{2}-1}{2}\label{B0}
\end{equation} 
and the asymptotic eigenvalues, $r_1=1/\sqrt{2}$ and $r_2=1-1/\sqrt{2}$ give the asymptotic entanglement level of  $\protect{\bar S_0\approx 0.872}$. 

We now return to the case of non-local initial conditions $\ket{\Psi_\pm}$ defined in eq.~(\ref{Psi_s}), for which 
\begin{equation}
|\tilde a_k(0)|^2=\left\{
\begin{array}{cr}
\cos^2 k&\mbox{ for }\ket{\Psi_+} \\
\sin^2 k &\mbox{ for }\ket{\Psi_-}.
\end{array} \right.
\end{equation} 
Since $\protect{\intk Q(k) \cos^2k = \intk Q(k) \sin^2k = \frac 12}$, we obtain from eq.~(\ref{b2-ave}), $A=C=1/2$ and the eigenvalues are also determined by $|B|$ alone.  Since $|\tilde a_k(0)|^2$ is still even, only the imaginary part of $R(k)$ makes a contribution. When inserted in eq.~(\ref{b-abs}), these initial conditions result in 
\begin{eqnarray}
B_+\equiv|\bar B(\Psi_+)|=&|\intk R(k)\,\cos^2k |&=\frac{(\sqrt{2}-1)^2}{2}\nonumber\\
B_-\equiv|\bar B(\Psi_-)|=&|\intk R(k)\,\sin^2k|& =\frac12-(\sqrt{2}-1)^2.\nonumber\\
&&\label{b1}
\end{eqnarray}
Note that these values are related by $\protect{B_0=\frac12\left(B_{-}+B_+ \right)}$. 
The exact eigenvalues are, 
\begin{equation}
r_1=2-\sqrt{2},\quad r_2=\sqrt{2}-1\qquad\mbox{for}~\ket{\Psi_+} \label{ev+}
\end{equation} 
and 
\begin{equation}
r_1=2(\sqrt{2}-1),\quad r_2=3-2\sqrt{2}\qquad\mbox{for}~\ket{\Psi_-},\label{ev-}
\end{equation}
and the corresponding asymptotic entanglements are, 
\begin{eqnarray}
\bar S_+&\equiv&\frac{1}{\sqrt{2}}-1-\log_2(\sqrt{2}-1)=0.97866\ldots\label{ent-pm}\\
\bar S_-&\equiv&-2(\sqrt{2}-1)\left( 1+\sqrt{2}\log_2(\sqrt{2}-1)\right) =0.66129\ldots, \nonumber
\end{eqnarray}
respectively. These exact values are coincident with those obtained numerically, see Fig.~\ref{fig:ent}. 
These are the maximum and minimum possible entanglement levels which can be obtained when starting in the position subspace ${\cal H}_1$ spanned by the kets $\ket{-1}$ and $\ket{1}$ with fixed initial coin. We show below that starting from a generic state in this subspace, all intermediate values of asymptotic entanglement are possible. 

\subsection*{Generic non-local initial state in ${\cal H}_1$}

Consider a generic ket in ${\cal H}_1$ as the initial condition for position. We keep the same initial coin $\ket{\chi}$ which leads to a symmetric evolution in the local case. Thus, we consider initial states of the form 
\begin{equation}
\ket{\Psi(\theta,\varphi)}=\left(\cos\theta\ket{-1}+e^{-i\varphi}\sin\theta\ket{1}\right)
\otimes\ket{\chi}.\label{Psi_gen}
\end{equation} 
The parameters $\protect{\theta\in[-\pi/2,\pi/2]}$ and $\protect{\varphi\in[-\pi,\pi]}$ are real angles. The initial amplitudes are $\protect{\tilde a_k(0)=\left(e^{ik}\cos\theta + e^{-i(k+\varphi)}\sin\theta\right)/\sqrt{2}}$, $\protect{\tilde b_k(0)=i\tilde a_k(0)}$ so, 
\begin{equation}
|\tilde a_k(0)|^2=|\tilde b_k(0)|^2=\frac12\left[1+\sin(2\theta)\cos(2k+\varphi)\right].\label{a2-init}
\end{equation} 
\begin{figure}
\includegraphics[scale=1]{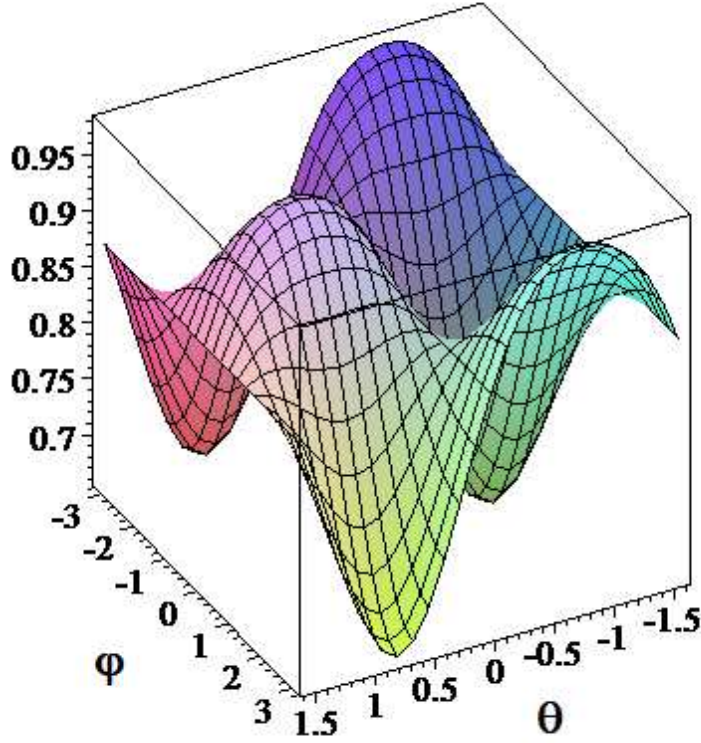}
\caption{Asymptotic entropy of entanglement $\bar S_E(\theta,\varphi)$ as a function of the initial state from eq.~(\ref{Psi_gen}). The entropy is calculated from the expressions for $\bar C$ and $|B|$ given in eqs.(\ref{C-gen1}) and (\ref{Bprime}).} 
\label{fig:sc4}
\end{figure}
We use eq.~(\ref{b2-ave}) to obtain,
\begin{eqnarray}
\bar A &=&\frac12 + \sin(2\theta)\sin(\varphi)\intk\, Q(k)\sin(2k)\nonumber\\
&=&\frac12 - B_+\sin(2\theta)\sin(\varphi). \label{C-gen1}
\end{eqnarray}

For arbitrary $\varphi$ and $\theta=0,\pm\pi/2$ (indicating localized initial positions) or for arbitrary $\theta$ and $\varphi=0,\pm\pi$ (indicating relative phases zero or $\pi$ between initial position eigenstates) we have $\protect{\bar A = \bar C = 1/2}$ and the asymptotic entropy is determined by the value of $|\bar B|$ alone.

The expression for $\bar B$ obtained from eq.~(\ref{b-abs}) is, 
\begin{equation}\label{Bprime}
 \bar B = -B_+\sin(2\theta)\sin(\varphi) -i(B_0 -B'\sin(2\theta)\sin(\varphi)) 
\end{equation}
with $B_0$ defined in eq.~(\ref{B0}) and 
\begin{equation}
B'\equiv\frac{ B_{-}-B_+}{2} =\frac{3\sqrt{2}-4}{2}.
\end{equation} 
\begin{figure}
\includegraphics[width=10cm]{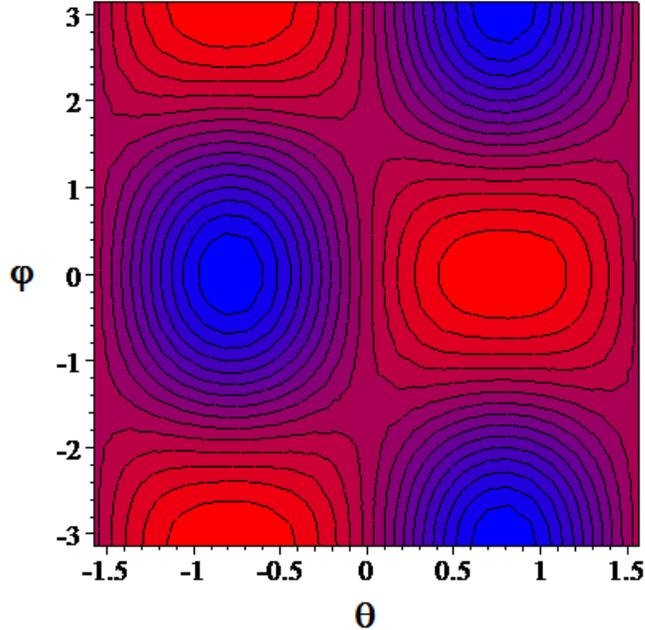}
\caption{Contour plot the surface shown in Fig.~\ref{fig:sc4}, using the eigenvalues from eq.~(\ref{ex-ev}). Red areas indicate maxima and blue areas, the minimum values.}
\label{fig:sc4-contour}
\end{figure}
The asymptotic eigenvalues $\bar r_{1,2}$ are obtained from eq.~(\ref{rho-ev}) as 
\begin{equation}
\bar r_{1,2}(\theta,\varphi)=\frac12 \pm \left[\left(B_0-B'\sin(2\theta)\cos(\varphi) \right)^2 + 2(B_+\sin(2\theta)\sin(\varphi))^2\right]^{1/2}.\label{ex-ev}
\end{equation}
The resulting asymptotic entanglement $\bar S_E(\theta,\varphi)$ is shown in Fig.~\ref{fig:sc4}. A contour plot of this surface, Fig.~\ref{fig:sc4-contour}, shows that there are only two maximum and two minimum points (for initial conditions  $\ket{\Psi_+}$ and $\ket{\Psi_-}$ or $(\theta,\varphi)=(\pi/4,0)$ and $(\pi/4,\pi)$) for which the asymptotic entanglements are $\protect{\bar S_+\approx 0.979}$ and $\protect{\bar S_-\approx 0.661}$, respectively. 

\begin{figure}
\includegraphics[scale=0.6]{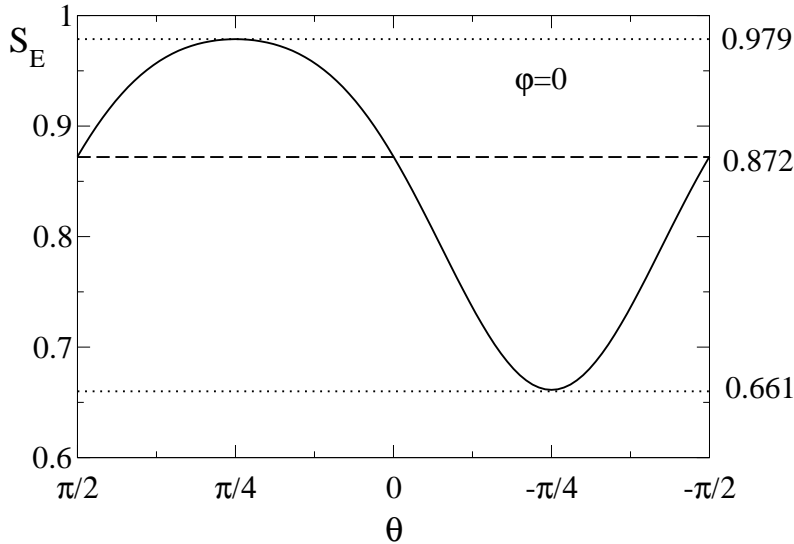}
\caption{Asymptotic entropy of entanglement $\bar S_E(\theta,\varphi=0)$ for initial states in ${\cal H}_1$ (see eq.~(\ref{Psi_gen})). Notice that the $\theta$ increases leftwise. This curve was generated using the eigenvalues given by eq.~(\ref{ex-ev}), with $\varphi=0$. The maximum and minimum values (dotted lines) can be calculated with arbitrary precision from the exact eigenvalues given in eqs.(\ref{ev+}) and eqs.(\ref{ev-}). The dashed line indicates the level of asymptotic entanglement associated to local initial conditions. } 
\label{fig:sc-theta}
\end{figure}

\subsection*{More general non-local initial states}

In order to further illustrate the effects that the non-locality in the initial condition can have on the asymptotic entanglement level, we consider an initial Gaussian wave packet with a characteristic spread $\sigma\gg 1$  in position space with the same coin state $\ket{\chi}$ as before. In this case, the Fourier transformed coefficients 
$\tilde a_k(0)\propto \sqrt{\sigma}\,e^{-k^2\sigma^2/2}$
correspond to a well-localized state in $k$ space with $\lim_{\sigma\rightarrow\infty} |\tilde a_k(0)|^2=2\pi\delta(k)$, where $\delta(k)$ is Dirac's delta function. In this limit, the elements of the asymptotic reduced density matrix can be trivially evaluated from eqs. (\ref{b2-ave}) and (\ref{b-abs})
\begin{eqnarray}
\bar A&=&\lim_{\sigma\rightarrow\infty}\intk Q(k)\,|\tilde a_k(0)|^2=Q(0)=1\nonumber\\
\bar B&=&\lim_{\sigma\rightarrow\infty}\intk R(k)\,|\tilde a_k(0)|^2=R(0)=0, \nonumber
\end{eqnarray} 
the eigenvalues are $r_1=1$ and $r_2=0$ and the corresponding asymptotic entropy of entanglement vanishes. 

This simple example shows that  for a particular uniform distribution in position space, a product state results in the long time limit. Of course, this is not true for arbitrary relative phases between the initial position eigenstates. However, it is clear that if more sites are initially occupied with appropriate relative phases, lower asymptotic entanglement levels may be obtained, until eventually a product state is reached.  A more detailed analysis of this example shows that for $\sigma\gg 1$, the smaller eigenvalue approaches zero as $(2\sigma)^{-4}$ and the asymptotic entropy of entanglement decays as 
$\protect{\bar S_E \sim \log_2\sigma /4\sigma^4+{\cal O}(\sigma^{-8})}$. Thus, for $\sigma\gtrsim 10$ the asymptotic entanglement is already quite small, $S_E\approx 10^{-4}$ . Thus, if many sites are initially occupied, the entanglement level at long times becomes  negligible. 

\section{Conclusions}
\label{sec:conc}

The long--time (asymptotic) entanglement properties of the Hadamard walk on the line are analytically investigated using the Fourier representation. The von Neumann entropy of the reduced density operator is used to quantify entanglement between the coin and position degrees of freedom. The fact that the evolution operator of a quantum walk is diagonal in $k$-space allows us to obtain clean, exact expressions for the asymptotic entropy of entanglement, $\bar S_E,$ for different classes of initial conditions. 

An expression for the exact value of asymptotic entanglement for localized initial states $\protect{\ket{0}\otimes\ket{\chi}}$, with arbitrary coin  $\protect{\ket{\chi}=\left( \cos\alpha\ket{R}+e^{i\beta}\sin\alpha\ket{L}\right)/\sqrt{2}}$, has been obtained analytically. 
This entanglement changes between full entanglement and a minimum of $\bar S_E=0.736\ldots$. For full entanglement, the convergence to the asymptotic value is significantly faster than in the case of minimum entanglement. Lower asymptotic entanglement levels may be obtained if non-local initial conditions are considered. 

We considered in detail the case of initial conditions in the position subspace spanned by $\ket{\pm 1}$, with fixed coin $(\ket{R}+i\ket{L}/\sqrt{2}$, and obtained an exact expression for the asymptotic entropy of entanglement. This expression shows that it varies smoothly between the extreme values $\bar S_-\approx 0.661$ and $\bar S_+\approx 0.979$.  As expected for this coin, the localized initial positions $\ket{\pm 1}$ have an intermediate entanglement of $\bar S_0=0.872\ldots$. In order to explore the effect of increasing non-locality, the asymptotic entanglement the case of an initial gaussian profile in position space, with characteristic spread $\sigma\gg 1$, is considered. For the particular phase relation considered, the resulting asymptotic entanglement decays fast with increasing initial non-locality. Thus, if many sites are initially occupied, a negligible entanglement level may be obtained at long times. 

The results presented in this work need to be extended to less simple systems. Most likely, quantum walks with either more particles, more dimensions or both will be required to be useful for algorithmic applications. The problem of entanglement in such systems is more involved. For example, in a quantum walk with two non-interacting particles there are four degrees of freedom and several kinds of entanglement  may coexist. Some initial work in this direction is presently under way.

\vskip 20mm
\textit{We thank Ms. Anette Gattner for pointing out an incorrect sign in the coefficient $b_3$ appearing in Appendix A in previous versions. This affected our previous conclusions for the case of localized initial conditions. We also thank Mostafa Annabestani for informing us of several corrections, including one which leads to the factor of 2 in eq.~(\ref{ex-ev}).} 

\vskip 10mm
\textit{We acknowledge support from PEDECIBA and PDT project 29/84. R.D. acknowledges financial support from  FAPERJ  (Brazil) and the Brazilian Millennium Institute  for Quantum Information--CNPq.}

\newpage
\appendix{\noindent\bf\large Appendix A: \\
Asymptotic entanglement level for localized initial conditions. }
\vskip1cm
In this Appendix we obtain the analytical expression for the dependence of the asymptotic entanglement level on the initial coin state for the case of initially localized position eigenstates. Let us consider as an initial condition a position eigenstate, which we take as as $x=0$ without loss of generality.  Let us also consider a generic coin 
\begin{equation}
\ket{\Psi(0)}=\ket{0}\otimes\frac{1}{\sqrt{2}}\left( \cos\alpha\ket{R}+e^{i\beta}\sin\alpha\ket{L}\right) 
\end{equation} 
where $\alpha, \beta$ are two real angles. The Fourier-transformed initial coefficients, eq.~(\ref{ft-a}), are
\begin{eqnarray}
\tilde a_0(0)=a_0(0)&=&\cos\alpha\nonumber\\
\tilde b_0(0)=b_0(0)&=&e^{i\beta}\sin\alpha
\end{eqnarray} 

From expressions (\ref{fg-exps}), for this class of initial conditions we obtain 
\begin{eqnarray}
|F_k|^2&=& \cos^2\alpha+|v_k|^2\sin^2\alpha+\sin(2\alpha)\,\real{u_k^*v_ke^{-i\beta}}\nonumber\\
&=&F_e(k)+F_o(k)\nonumber\\
|G_k|^2&=&\cos^2\alpha+|w_k|^2\sin^2\alpha+\sin(2\alpha)\,\real{u_k^*w_ke^{-i\beta}}.\nonumber\\
&=&G_e(k)+G_o(k) \label{fg-exps-app}
\end{eqnarray} 
where the even and odd parts are the real functions
\begin{eqnarray}
F_e(k)&\equiv& \cos^2\alpha+|v_k|^2\sin^2\alpha  \nonumber\\
&&\qquad\qquad -\left(1-\sqrt{2}\cos(\omega_k-k) \right)\sin(2\alpha)\cos\beta \nonumber\\
F_o(k)&\equiv&-\sqrt{2}\sin(\omega_k-k)\sin(2\alpha)\sin\beta \nonumber\\
G_e(k)&\equiv& \cos^2\alpha+|w_k|^2\sin^2\alpha  \label{fg-evodd}\\
&&\qquad\qquad-\left(1+\sqrt{2}\cos(\omega_k+k) \right)\sin(2\alpha)\cos\beta \nonumber\\
G_o(k)&\equiv&-\sqrt{2}\sin(\omega_k+k)\sin(2\alpha)\sin\beta. \nonumber
\end{eqnarray} 

The eigenvalues (\ref{rho-ev}) of the reduced density operator, which determine the entanglement level, depend only on its determinant 
\begin{equation}
\Delta = AC-|B|^2=C-(C^2+|B|^2).\label{det}
\end{equation} 
The asymptotic entanglement will be independent of the initial coin state if $\bar\Delta$ does not depend on $\alpha$ or $\beta$. 
From the general asymptotic expressions (\ref{ab11}) which are valid in the long-time limit for arbitrary initial conditions, we obtain 
\begin{eqnarray}
\bar C&=&c_1+(c_2-c_1)\sin^2\alpha+c_3\sin(2\alpha)\cos\beta \nonumber\\
\bar B&=&b_1+(b_2-b_1)\sin^2\alpha+\sin(2\alpha)\left(b_3\cos\beta +ib_4\sin\beta \right) \nonumber\\
\label{BC}
\end{eqnarray}
where  the real coefficients $c_j,b_j$ are explicitly
\begin{eqnarray}
c_1&=&\intk\left(\alpha_k^4+\beta_k^4 \right)=1-\frac{\sqrt{2}}{4}\nonumber\\
c_2&=&\intk\left(\alpha_k^4|v_k|^2+\beta_k^4|w_k|^2 \right) =\frac{\sqrt{2}}{4}\nonumber\\
b_1&=&c_3=-\intk\left[\alpha_k^4\left(1-\sqrt{2}\cos(\omega_k-k) \right) +\right. \nonumber\\
&&\qquad\qquad\left.\beta_k^4\left(1+\sqrt{2}\cos(\omega_k+k) \right)  \right]=\frac12-\frac{\sqrt{2}}{4}\nonumber\\
b_2&=&-\intk\left[\alpha_k^4|v_k|^2\left(1-\sqrt{2}\cos(\omega_k-k) \right) +\right.\label{coef1}\\
&&\qquad\qquad\left.\beta_k^4|w_k|^2\left(1+\sqrt{2}\cos(\omega_k+k) \right)  \right]=\frac{\sqrt{2}}{4}-\frac12\nonumber\\
b_3&=&\intk \left[\alpha_k^4\left(1-\sqrt{2}\cos(\omega_k-k)^2 \right)^2+\right.\nonumber\\
&&\qquad\qquad\left.\beta_k^4\left(1+\sqrt{2}\cos(\omega_k+k) \right)^2\right]=\frac12-\frac{\sqrt{2}}{4}\nonumber\\
b_4&=&2\intk\left[\alpha_k^4\sin^2(\omega_k-k) +\right.\nonumber\\
&&\qquad\qquad\qquad\qquad\left.\beta_k^4 \sin^2(\omega_k+k)\right]=\frac{\sqrt{2}-1}{2}.\nonumber
\end{eqnarray} 
These coefficients can be conveniently expressed in terms of one of them, i.e $b_1$, 
since they satisfy the relations
\begin{equation}
\begin{array}{ccc}
c_1=b_1+\frac12 ,\qquad&c_2-c_1=-2b_1,&\qquad c_3=b1,\nonumber\\
b_2=-b_1,&b_3=b_1,&b_4=\sqrt{2}\,b_1.\nonumber
\end{array}
\end{equation} 

Thus, eqs.~(\ref{BC})  can be expressed as
\begin{eqnarray}
\bar C&=&b_1\left[3+\sqrt{2}-2\sin^2\alpha+\sin(2\alpha)\cos\beta \right]\nonumber\\
\bar B&=&b_1\left[1-2\sin^2\alpha+\sin(2\alpha)\left(\cos\beta+i\sqrt{2}\sin\beta \right)  \right]\nonumber\\.
\end{eqnarray}  
and  the exact determinant is,,  
\begin{equation}
\Delta=\Delta_0-2b_1^2\,\cos\beta\,\sin(4\alpha) \label{Delta}
\end{equation}
where
\begin{equation}
\Delta_0=c_1(1-c_1)-b_1^2=\frac{\sqrt{2}-1}{2}.\label{Delta0}
\end{equation} 
These expressions can be used to calculate the exact asymptotic value for the entropy of entanglement for local initial states with arbitrary coins.

\bibliography{qwnloc}
\bibliographystyle{h-physrev} 

\end{document}